\documentclass[letterpaper,journal]{IEEEtran}
\usepackage{graphicx}
\usepackage{float}
\usepackage{amsmath}
\usepackage{epstopdf}
\usepackage{cases}
\usepackage{color}
\usepackage{xcolor}
\usepackage{algorithm}
\usepackage{algpseudocode}
\usepackage{amsmath}
\usepackage{amssymb}
\usepackage{subcaption}
\usepackage{soul}
\usepackage{hyperref}
\makeatletter
\renewcommand{\maketag@@@}[1]{\hbox{\m@th\normalsize\normalfont#1}}%
\makeatother
\usepackage{stfloats}
\usepackage{cite}
\usepackage{makecell}
\usepackage{multirow}
\usepackage{ulem}

\usepackage{tabularx}  
\usepackage{booktabs}  

\ifCLASSINFOpdf
\else
\fi

\usepackage{mathtools}

\usepackage{acro}[=v3]
\acsetup{
	use-id-as-short=true,
	make-links=false,
	link-only-first=true,
}
\makeatletter
\DeclareAcronym{IoT}{
	long={internet of things},
}
\DeclareAcronym{AI}{
	long={artificial intelligence},
}
\DeclareAcronym{LEO}{
	long={low Earth orbit},
}
\DeclareAcronym{3G}{
	long={the third-generation},
}
\DeclareAcronym{4G}{
	long={the fourth-generation},
}
\DeclareAcronym{5G}{
	long={the fifth-generation},
}
\DeclareAcronym{6G}{
	long={The sixth-generation},
}
\DeclareAcronym{NR}{
	long={new radio},
}
\DeclareAcronym{BS}{
	long={base stations},
}
\DeclareAcronym{SGCE}{
	long={space-ground coordinate entity},
}
\DeclareAcronym{SCE}{
	long={satellite coordinate entity},
}
\DeclareAcronym{GCE}{
	long={ground coordinate entity},
}
\DeclareAcronym{3GPP}{
	long={the 3rd generation partnership project},
}
\DeclareAcronym{NTN}{
	long={non-terrestrial networks},
}
\DeclareAcronym{DHTS}{
	long={direct handset-to-satellite},
}
\DeclareAcronym{SGCN}{
	long={space-ground collaborative network},
}
\DeclareAcronym{SGIN}{
	long={space-ground integrated network},
}
\DeclareAcronym{eMBB}{
	long={enhanced mobile broadband},
}
\DeclareAcronym{mMTC}{
	long={massive machine-type communications},
}
\DeclareAcronym{URLLC}{
	long={ultra-reliable low-latency communications},
}
\DeclareAcronym{HAPS}{
	long={high-altitude platform stations},
}
\DeclareAcronym{TN}{
	long={terrestrial networks},
}
\DeclareAcronym{SAGIN}{
	long={space-air-ground integrated network},
}
\DeclareAcronym{NF}{
	long={network functions},
}
\DeclareAcronym{NE}{
	long={network element},
}
\DeclareAcronym{QoS}{
	long={quality of service},
}
\DeclareAcronym{QoE}{
	long={quality of experience},
}
\DeclareAcronym{TSN}{
	long={time-sensitive networking},
}
\DeclareAcronym{MIMO}{
	long={multiple-input multiple-output},
}
\DeclareAcronym{ELAA}{
	long={extremely large aperture array},
}
\DeclareAcronym{THz}{
	long={terahertz frequencies},
}
\DeclareAcronym{mmWave}{
	long={millimeter-wave},
}
\DeclareAcronym{Gbps}{
	long={Gbits per second},
}
\DeclareAcronym{ISL}{
	long={inter‑satellite links},
}
\DeclareAcronym{D2D}{
	long={device-to-device},
}
\DeclareAcronym{RRC}{
	long={radio resource control},
}
\DeclareAcronym{IoV}{
	long={internet of vehicles},
}
\DeclareAcronym{RTT}{
	long={long round-trip time},
}
\DeclareAcronym{XR}{
	long={extended reality},
}
\DeclareAcronym{LPWAN}{
	long={low-power wide-area networks},
}
\DeclareAcronym{gNB}{
	long={next generation node B},
}
\DeclareAcronym{S-gNB}{
	long={satellite gNBs},
}
\DeclareAcronym{G-gNB}{
	long={ground gNBs},
}
\DeclareAcronym{UE}{
	long={user equipment},
}
\DeclareAcronym{SGCAN}{
	long={space-ground cooperative access Network},
}
\DeclareAcronym{CU}{
	long={centralized unit},
}
\DeclareAcronym{DU}{
	long={distributed unit},
}
\DeclareAcronym{RU}{
	long={radio units},
}
\DeclareAcronym{MAC}{
	long={medium access control},
}
\DeclareAcronym{CoMP}{
	long={coordinated multiple points},
}
\DeclareAcronym{SIR}{
	long={signal-to-interference ratio},
}
\DeclareAcronym{ISCC}{
	long={Integrated sensing, communication and computing},
}
\DeclareAcronym{SDN}{
	long={software-defined networking},
}
\DeclareAcronym{CN}{
	long={core network},
}
\DeclareAcronym{E2E}{
	long={end-to-end},
}
\DeclareAcronym{SWaP}{
	long={size, weight and power},
}
\DeclareAcronym{PRB}{
	long={physical resource blocks},
}
\DeclareAcronym{SNR}{
	long={signal-to-noise ratio},
}
\DeclareAcronym{STIN}{
	long={satellite-terrestrial integrated network},
}
\DeclareAcronym{O-RAN}{
	long={open radio access network alliance},
}
\DeclareAcronym{IA}{
	long={interference alignment},
}
\DeclareAcronym{DC}{
	long={dual-connectivity},
}
\DeclareAcronym{RL}{
	long={reinforcement learning},
}

\immediate\write18{texcount -inc -sum main.tex > wordcount.txt}

\begin{document}
\title{A Hierarchical Space-Ground Collaborative Network: Architecture Design and Deployment}
\author{
        Yunxin Geng,~\IEEEmembership{Graduate Student Member,~IEEE,}
        Shaopeng Feng, Xuekun Hao,
        Zijian Guo, \\
        Shengfeng Wang, 
        Rui Ding,
        Kan Zheng,~\IEEEmembership{Fellow,~IEEE,}
        Zhiqing Wei,~\IEEEmembership{Member,~IEEE,} \\
        Weiwei Jiang,~\IEEEmembership{Senior Member,~IEEE,} 
        and Zhiyong Feng,~\IEEEmembership{Senior Member,~IEEE}
        
\thanks{
Yunxin Geng, Shaopeng Feng (corresponding author), Zijian Guo and Shengfeng Wang are with State Key Laboratory of Satellite Network, Shanghai Satellite Network Research Institute Co., Ltd., and Shanghai Key Laboratory of Satellite Network, Shanghai, China.
Yunxin Geng is also with Beijing University of Posts and Telecommunications, Beijing, China.

Xuekun Hao and Rui Ding are with China Satellite Network Group Company Ltd., Beijing, China.
Xuekun Hao is also with State Key Laboratory of Satellite Network, Shanghai, China.

Kan Zheng is with Ningbo University, Ningbo, China.

Zhiqing Wei, Weiwei Jiang and Zhiyong Feng are with Beijing University of Posts and Telecommunications, China.
Weiwei Jiang is also with Anhui Provincial Key Laboratory of Multimodal Cognitive Computation, Anhui University, Hefei, China.
}
\thanks{Corresponding author: Shaopeng Feng (\href{mailto: fspsd@sina.com}{fspsd@sina.com}).}
}

\maketitle

\begin{abstract} 
The sixth-generation (6G) landscape demands a profound architectural evolution toward a unified, collaborative ecosystem that integrates satellite and terrestrial segments to provide ubiquitous global connectivity.
Nevertheless, conventional independent management paradigms are increasingly challenged by the profound spatiotemporal nonstationarity of low-Earth orbit (LEO) channels and the volatile demands of massive heterogeneous services.
To address these challenges, this study focuses on the architectural design of space-ground integrated networks (SGINs) and investigates elastic deployment strategies tailored to heterogeneous 6G service requirements. We first review the development trends in SGIN and then summarize typical application scenarios into two categories, along with their associated challenges and constraints. 
After that, a hierarchical space-ground collaborative network (SGCN) architecture is proposed to support achieving larger network capability, extensive continuous coverage, and enhanced spectral efficiency, which is evaluated through three representative scenarios with progressive baselines. Subsequently, a service-aware deployment strategy is introduced to further establish the engineering feasibility of the proposed architecture. 
\end{abstract}

\section{Introduction}
\label{sec:introduction}
\ac{6G} networks aim to establish ubiquitous global connectivity, realizing the vision of ``connected intelligence'' through seamless service provisioning in remote regions~\cite{xiao2024guest}.
This systemic transformation is propelled by the escalating demand for reliable infrastructure capable of supporting high-speed data transmission across diverse environments.
However, \ac{TN}, constrained by deployment costs and geography, covers only a fraction of the Earth's surface.
In this context, \ac{SGIN} has emerged as a pivotal solution. 
By establishing a unified framework that effectively integrates \ac{TN} and \ac{NTN},
\ac{SGIN} promises to offer enhanced scalability, reliability and flexibility, enabling seamless communication in areas where traditional ground infrastructure is either non-existent or insufficient.

Although the \ac{3GPP} standard initially introduced the concepts of ``transparent forwarding'' and ``regenerative'' architectures in Release 16~\cite{3gpp.38.821} and has since evolved its standardization efforts, current satellite and terrestrial networks operate largely as isolated parts. 
This lack of deep collaboration introduces two fundamental categories of challenges:
\begin{itemize}
    \item \textbf{Multi-dimensional spatiotemporal scale mismatch: }
    The mismatch originates from the fundamental heterogeneity and lack of unified interfaces between \ac{LEO} satellites and terrestrial \ac{gNB}.
    Moreover, protocol stack incompatibility between satellite systems and 
    \ac{5G} \ac{NR}, as well as the extreme topological dynamics of \ac{LEO} satellites, respectively lead to excessive conversion overhead and significant control-plane decision-making delay~\cite{zhang2025Direct}.

    \item \textbf{High-dynamic environment constraints and service assurance:}
    The requirement for repeated user authentication during cross-network mobility results in significant service interruption. 
    Furthermore, satellite payloads are strictly constrained by \ac{SWaP} limitations, which hinder the support of computation-intensive services. 
    Finally, an uneven user distribution can overload constrained feeder links with bursty traffic.
\end{itemize}

To address the aforementioned challenges, substantial research efforts have been dedicated to satellite-terrestrial integration from two primary perspectives: architectural design and functional optimization~\cite{Xiao2024space}. 
From an architectural perspective, several studies have explored the fusion of space and ground segments using evolving network paradigms.
For instance, the \ac{SDN} framework has been leveraged to integrate space-terrestrial components, aiming to provide improved \ac{QoS} for global multimedia services~\cite{Bi2019software}. 
Other frameworks focus on the general vision of \ac{SAGIN} to deliver ubiquitous coverage and predictable low-latency connectivity~\cite{zhang20256G}. 
However, a common limitation of these studies is that they often treat satellite and terrestrial networks as separate administrative domains, which lack a unified control plane capable of orchestrating resources across both segments in a truly integrated manner.
Regarding specific functional optimizations, the existing literature has proposed various solutions for resource allocation, routing, and mobility management.
Cloud-edge aggregated \ac{AI} architectures have been designed to enhance network throughput~\cite{Xu2023ai}, whereas big data-driven routing frameworks and spectrum virtualization technologies have been introduced to handle heterogeneous traffic demands~\cite{wang2019tensor,Xiao2024space}.
Furthermore, sophisticated handover strategies have been developed to maintain service continuity in \ac{NTN}~\cite{zhu2022integrated,wang2023seamless}.
Critically, most of these efforts have addressed technical issues in isolation. 
They focus on single-dimension metrics without providing a unified architectural framework that can dynamically adapt \ac{NE} deployment to the primary barriers in \ac{SGIN}.

Hence, a hierarchical and service-oriented \ac{SGCN} architecture is proposed, 
This can reconcile the conflict between management simplicity and system resilience.
The main contributions of this study are summarized as follows.
\begin{itemize}
    \item \textbf{\ac{SGIN} evolution trends and scenario analysis: }
    We first introduce the core concepts, principal challenges, and evolution path of \ac{SGIN}.
    We then summarize the typical application scenarios and highlight some requirements.
    \item \textbf{Hierarchical \ac{SGCN} architecture with standards mapping:} 
    A novel hierarchical SGCN architecture is proposed that decouples global orchestration from local real-time control, enabling the efficient management of heterogeneous nodes and optimal resource allocation.
    We explicitly map each new collaborative interface in \ac{SGCN} to \ac{3GPP} standardized counterparts.
    \item \textbf{Representative scenario simulations:} We evaluated the proposed \ac{SGCN} under three scenarios with four baselines, ranging from no cooperation and industry-standard approaches to centralized gateway coordination. The results evidently show performance gains in network capacity, continuous coverage, and spectral efficiency.
    \item \textbf{Flexible \ac{NE} deployment strategy with quantified \ac{SWaP}:}
    By decoupling \ac{NE}s from physical nodes, the proposed architecture supports flexible deployment strategies
    tailored to the specific service requirements.
    Furthermore, the service-aware deployment decision framework and quantified \ac{SWaP} analysis establish engineering feasibility.
\end{itemize}

\begin{figure*}[tb]
    \centering
    \includegraphics[width=0.95\textwidth]{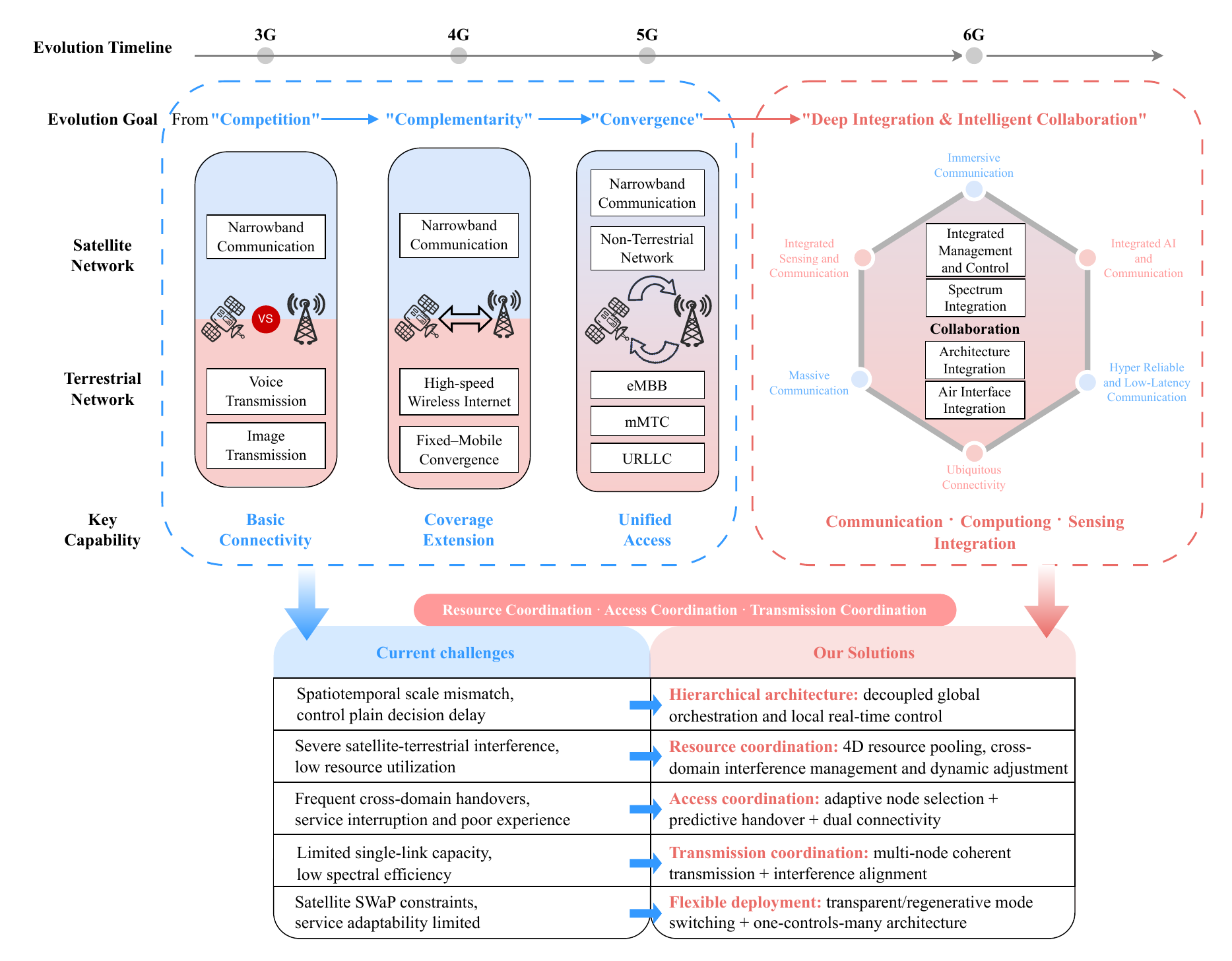}
    \caption{The evolution trends of satellite-terrestrial integrated network.}
    \label{fig:evolution}
\end{figure*}
\section{Foundation of Satellite-Terrestrial Collaborative Network}
This section first outlines the development trends in satellite-terrestrial integrated networks. 
We then categorized the application scenarios of \ac{SGIN} into two primary types.
Subsequently, the key requirements underpinning these application scenarios are discussed.

\subsection{Evolution Trends of Satellite-Terrestrial Integrated Network}
For a long time, satellite communications and terrestrial mobile communications have developed in two largely independent ways. 
Specifically, \ac{NTN}, serving as a vital complement to \ac{TN}, utilizes platforms such as satellites to enable seamless global coverage, emergency communications, and integrated air-space-ground-sea services. 

As illustrated in Fig.~\ref{fig:evolution}, the trend of satellite-terrestrial integration has evolved through the stages of ``competition in \ac{3G}, complementarity in \ac{4G}, and convergence in \ac{5G}'', and is progressively moving towards the ``integration'' phase in \ac{6G}. 
From Release 15, \ac{3GPP} has gradually incorporated \ac{NTN} into its standardization framework, initiating discussions on network architectures, application scenarios, and technical solutions~\cite{3gpp.38.811}.
Release 17 marked the first standardization of \ac{5G} \ac{NR}-based \ac{NTN} with transparent payloads~\cite{3gpp.38.300}
and since Release 19, the focus has shifted to standardizing regenerative architectures and exploring flexible onboard functional splits~\cite{3gpp.38.821}.

\ac{6G} has laid out the development roadmap for a deeply integrated space-terrestrial network, which achieves convergence across four key dimensions: architecture, air interface, spectrum, as well as management and control.
The ultimate objective is to propose standardized solutions for \ac{6G}, realize intrinsic system-level integration, and finally achieve global seamless coverage.

\subsection{Typical Application Scenarios and Key Requirements}
As summarized in Table~\ref{tab:scenarios}, we categorize the application scenarios broadly into two primary classes.

\subsubsection{Classification of Collaborative Scenarios and Performance Bottlenecks}

\textbf{Satellite-terrestrial complementary coverage} addresses the ``connectivity gap'' where terrestrial infrastructure is absent or limited.
This class can be further subdivided into non-overlapping and overlapping coverage.

\begin{itemize}
    \item \textit{Non-overlapping coverage:}
    This category encompasses scenarios such as communication in remote mountainous territories 
    and massive \ac{IoT} deployments in uninhabited areas. 
    Without collaboration, these regions suffer from fragmented connectivity voids and spectral fragmentation because of static frequency partitioning. 
    While system-level integration overcomes these barriers, it establishes an uninterrupted service fabric across global coordinates and maximizes the efficiency of orthogonal frequency blocks through dynamic sharing.
    \item \textit{Overlapping coverage:}
    Typical cases include high-altitude civil aviation and terrestrial coverage holes in urban canyons. 
    In a non-collaborative state, users encounter ambiguous association policies, severe cross-layer co-channel interference, and fluctuating \ac{E2E} performance.
    Collaborative mechanisms facilitate resilient link association and interference nulling, thereby stabilizing the throughput and ensuring link robustness.
\end{itemize}

\textbf{Satellite-terrestrial superposition enhancement} provides a performance ``booster'' where terrestrial coverage exists but is insufficient for peak demands. 
This is subdivided into capacity and service continuity enhancement.
\begin{itemize}
    \item \textit{Capacity enhancement:}
    This category of scenarios targets situations such as traffic tidal effects 
    and emergency rescue missions, respectively.
    Non-collaborative systems struggle with access stratum congestion and bottlenecks in feeder-link bandwidth during bursts. 
    Deep collaboration enables the elastic provisioning of supplemental bandwidth and effectively buffers burst traffic through opportunistic offloading.
    \item \textit{Service continuity enhancement:}
    Applied to high-mobility use cases like \ac{IoV} and cross-border transportation. 
    Isolated systems suffer from frequent connection drops and \ac{RRC} re-establishment delays during cross-network transitions.
    Collaborative orchestration ensures transparent mobility management with negligible handover latency and maintains persistent session states even at high orbital velocities.
\end{itemize}

\begin{table*}[tb]
  \centering
  \caption{Summary of typical application scenarios for satellite-terrestrial collaborative network}
  \label{tab:scenarios}
  \renewcommand{\arraystretch}{1.5}
  \resizebox{0.95\textwidth}{!}{%
  \begin{tabular}{c|l|l|l|l|l}
    \toprule
    \textbf{Category} &
    \textbf{Characteristics} &
    \textbf{Scenario type} &
    \textbf{Typical use cases} &
    \textbf{Challenges without collaboration} &
    \makecell[l]{\textbf{Capabilities with space-ground} \\ \textbf{collaboration}} \\
    \midrule

    \multirow{2}{*}[-4ex]{%
      \makecell[c]{\textbf{Satellite-} \\ 
      \textbf{terrestrial} \\ 
      \textbf{complementary} \\ 
      \textbf{coverage}}}
      
    & \multirow{2}{*}[0.5ex]{%
        \makecell[l]{
        i) Global and all‑time continuous \\coverage \\
        ii) Support for highly mobile terminals \\
        iii) Dispersed users with diverse \\service requirements \\
        iv) Massive \ac{IoT} connectivity and \\scheduling demands \\
        v) High user‑experienced data rate
        }}

    & \begin{tabular}[c]{@{}l@{}}
        Non-overlapping \\ coverage
          \end{tabular}
    
    & \begin{tabular}[c]{@{}l@{}}
        Remote areas, \\
        maritime communications, \\
        massive \ac{IoT}
      \end{tabular}
      
    & \begin{tabular}[c]{@{}l@{}}
        Large coverage gaps,\\
        low spectrum utilization
      \end{tabular}
      
    & \begin{tabular}[c]{@{}l@{}}
        Enabling global wide‑area and all‑time \\continuous connectivity, \\
        significantly enhancing coverage range \\and spectrum utilization efficiency
      \end{tabular} \\
    \cline{3-6}

    & &
    \begin{tabular}[c]{@{}l@{}}
            Overlapping\\coverage
          \end{tabular}

    & \begin{tabular}[c]{@{}l@{}}
        Aeronautical communications,\\
        terrestrial coverage holes
      \end{tabular}
    & \begin{tabular}[c]{@{}l@{}}
        Complex access selection, \\
        satellite-terrestrial interference,\\
        unstable \ac{E2E} performance
      \end{tabular}
    & \begin{tabular}[c]{@{}l@{}}
        Enabling collaborative access and \\link selection, \\
        enhancing user access reliability \\and service stability
      \end{tabular} \\
    \midrule

    \multirow{2}{*}[-5ex]{%
      \makecell[c]{\textbf{Satellite-} \\ 
      \textbf{terrestrial} \\ 
      \textbf{superposition} \\ 
      \textbf{enhancement}}}

    & \multirow{2}{*}[-2.5ex]{%
        \makecell[l]{
        i) Support for multi-connectivity \\and multi-band \\
        ii) Unified orchestration of satellite-\\terrestrial resources \\
        iii) High security and high \\reliability \\
        }}
        
    & \begin{tabular}[c]{@{}l@{}}
        Capacity\\enhancement
      \end{tabular}

    & \begin{tabular}[c]{@{}l@{}}
        Traffic surges, \\
        emergency scenarios, \\
        hotspot areas
      \end{tabular}
      
    & \begin{tabular}[c]{@{}l@{}}
        Challenge of massive access, \\
        limited transmission capacity
      \end{tabular}
      
    & \begin{tabular}[c]{@{}l@{}}
        Enabling on‑demand capacity expansion \\
        to mitigate hotspot and burst traffic impact
      \end{tabular} \\
    \cline{3-6}

    & &
    \begin{tabular}[c]{@{}l@{}}
        Service continuity\\enhancement
      \end{tabular}

    & \begin{tabular}[c]{@{}l@{}}
        \ac{IoV}, \\
        cross-border railways
      \end{tabular}
    & \begin{tabular}[c]{@{}l@{}}
        Frequent handovers, \\
        service interruptions, \\
        difficulty of zero-perception access
      \end{tabular}
    & \begin{tabular}[c]{@{}l@{}}
        Enabling seamless handover and \\persistent connectivity, \\
        ensuring highly continuous, reliable, \\and secure services
      \end{tabular} \\
    \bottomrule
  \end{tabular}}
\end{table*}

\subsubsection{Scenario‑Driven Requirements for \ac{SGIN} Architectures}
The primary objective of \ac{SGIN} lies in eliminating structural and operational fragmentation between satellite and terrestrial segments, achieving
a globally orchestrated, unified network.
The implementation of such an integrated architecture necessitates the fulfillment of the following four fundamental requirements, which enable the network's robust scenario adaptability and dynamic responsiveness to fluctuating traffic demands.

\begin{itemize}
    \item \textbf{Coordination between \ac{LEO} and terrestrial Segments:}
    The architecture must adapt to differentiated \ac{QoS} demands, ranging from high-bandwidth consumer-oriented video applications to ultra-reliable business-oriented industrial control.
    \item \textbf{Compatibility with hybrid transparent and regenerative payloads:}
    The unified system must accommodate both bent-pipe forwarding for rapid, low-cost deployment and regenerative onboard processing to support low-latency secure private networking and local switching.
    \item \textbf{Distributed and modular functional configuration:}
    \ac{NE} must be flexibly split and deployed across satellites, ground gateways and cloud centers to optimize resource allocation based on real-time traffic distributions and hardware resource constraints.
    \item \textbf{High-low frequency coordination:}
    To ensure imperceptible service transitions, the architecture must coordinate low-frequency bands (e.g., L/S) for wide-area coverage with high-frequency bands (e.g., Ka/Ku) for data-intensive broadband services, 
    while maintaining precise cross-domain interference control.
\end{itemize}

\section{Design of Space-Ground Collaborative Network Architecture}
\label{sec3}
The preceding analysis underscores that legacy administrative silos in network management are incapable of handling the extreme dynamics and spatiotemporal mismatches inherent to \ac{SGIN}.
Motivated by these challenges, it is imperative to develop a hierarchical service-oriented architecture optimized for unified resource orchestration, high-capacity rapid access, and spectrally efficient transmission.

\begin{figure*}[tb]
    \centering
    \includegraphics[width=0.96\textwidth]{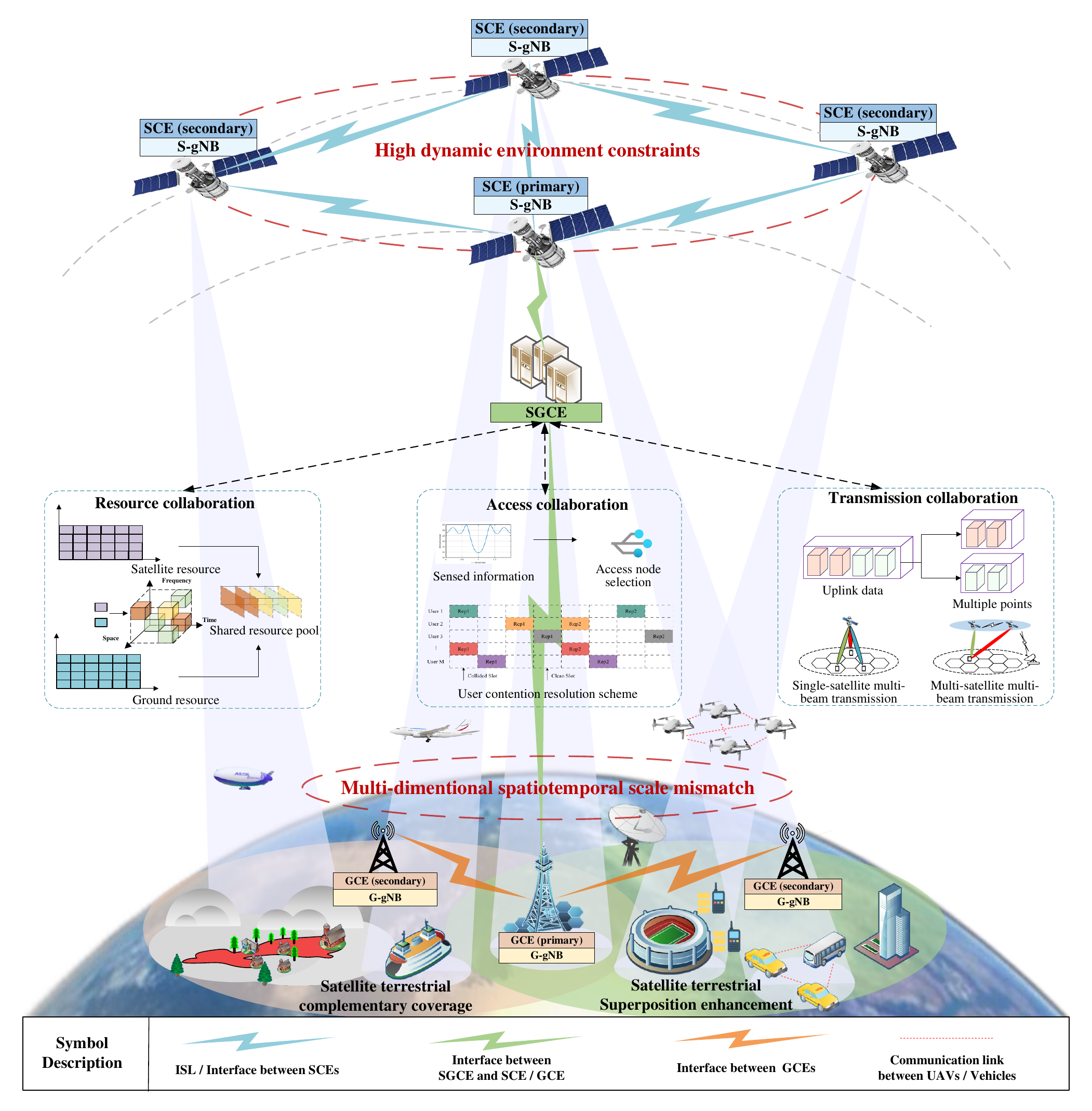}
    \caption{Design of the proposed hierarchical \ac{SGCN} architecture.}
    \label{fig:architecture}
\end{figure*}

\subsection{Hierarchical \ac{SGCN} Architecture}
Satellite‑terrestrial integration extends beyond basic connectivity, which demands a well-redesigned network architecture capable of resolving the inherent challenges like decision‑making delays and multi‑dimensional attribute asymmetries induced by satellite mobility.
However, legacy architectural paradigms are inherently constrained by the trade-off between orchestration simplicity and systemic resilience. 
Specifically, centralized architectures facilitate unified control but suffer from single points of failure and scalability bottlenecks. 
Conversely, distributed frameworks offer robustness but often succumb to excessive coordination overheads and decision locality.
To reconcile these conflicting requirements, we propose a hierarchical service-oriented architecture, e.g. \ac{SGCN}, which is illustrated in Fig.~\ref{fig:architecture}. By offloading real-time processing to distributed elements, such as \ac{SCE}/\ac{GCE}, we enhance local reliability and scalability. 
Simultaneously, a centralized global coordination entity, e.g. \ac{SGCE}, 
ensures policy consistency and cross-domain manageability.
Deployed on the terrestrial cloud or \ac{CN}, \ac{SGCE} orchestrates cross-domain resource allocation, transmission scheduling, and access management, unifying control across satellite and terrestrial planes.
Within either the satellite plane or the terrestrial plane, there exist \ac{GCE} and \ac{SCE}. 
\ac{GCE} is responsible for coordinating resources among multiple \ac{G-gNB}, while \ac{SCE} coordinates resources among multiple \ac{S-gNB}, thereby enabling unified access and transmission within the terrestrial and satellite segments respectively. 
Furthermore, \ac{SCE}/\ac{GCE} coordinates one or multiple \ac{S-gNB}s/\ac{G-gNB}s, undertaking localized collaborative management and control functions. 
\ac{SCE}s/\ac{GCE}s can form multi‑level cooperative structures, establishing primary \ac{SCE}s/\ac{GCE}s and secondary \ac{SCE}s/\ac{GCE}s to enable coordination across larger‑scale \ac{S-gNB}s or \ac{G-gNB}s clusters. Thus, the architecture can also be regarded as the following three virtual functional tiers:
\begin{itemize}
    \item \textbf{Global Management Tier:}
    \ac{SGCE} maintains a comprehensive view of the constellation ephemeris and terrestrial load status.
    In functional implementation, information exchange between \ac{SGCE}, \ac{GCE}, and \ac{SCE} enables space‑ground spectrum situation awareness and performs functions such as space‑ground resource allocation, interference management, load balancing, and adaptive node access.
    \item \textbf{System Coordination Tier:} \ac{SCE} and \ac{GCE} manage localized domains (e.g., satellite beams or terrestrial clusters) and perform rapid coordination tasks under the policy constraints set by \ac{SGCE}. 
    Furthermore, the primary \ac{SCE}/\ac{GCE} performs resource orchestration and management based on the network status and service demands reported by the \ac{gNB}s or secondary \ac{SCE}s/\ac{GCE}s, enabling coordinated decision‑making for multi‑satellite networking under its regional jurisdiction.
    The secondary \ac{SCE}/\ac{GCE} may assist the primary \ac{SCE}/\ac{GCE} in carrying out collaborative management decisions within the region, or it may conduct distributed decision-making by coordinating internally within \ac{gNB}s.
    \item \textbf{Node Execution Tier:}
    This tier comprises the physical access nodes of \ac{CU} and \ac{DU} that execute specific physical and \ac{MAC} layer instructions, such as beamforming weights and resource block allocations.
\end{itemize}

Compared with purely centralized and distributed alternatives, the three-tier \ac{SGCN} offers two decisive advantages. 
On the one hand, it confines coordination to intra-domain neighborhoods under each \ac{SCE}/\ac{GCE}, restraining the signaling explosion of fully centralized pairwise exchanges while retaining the global policy consistency that distributed designs cannot support. 
In contrast, the primary-secondary structure can strengthen system resilience: during network partitions, each region operates autonomously under its secondary \ac{SCE}/\ac{GCE} until the \ac{SGCE} resolves conflicts upon healing.

\subsection{Collaborative \ac{NE} Functions and Signaling Interfaces}
The focus of the \ac{SGCN} architecture lies in designing collaborative \ac{NE}s' functions and their mutual interfaces, which are essential for achieving the unified orchestration of communication, computing, and other assets across space and ground segments.





\subsubsection{\ac{SCE} and \ac{GCE}}
\ac{SCE} focuses on on-orbit management, including \ac{ISL} routing, real-time beam hopping and multi-satellite synchronization for \ac{CoMP}, while \ac{GCE} is tasked with orchestrating load balancing in the terrestrial domain, performing real-time interference monitoring for satellite-to-ground downlinks and executing adaptive traffic steering to mitigate localized \ac{TN} congestion.
\begin{itemize}
    \item \ac{SCE}/\ac{GCE} performs resource orchestration and management based on status and service‑demand information reported by \ac{S-gNB}/\ac{G-gNB}, enabling coordinated decision‑making for multi‑satellite/multi-\ac{TN} networking within its jurisdictional region.
    \item \ac{SCE}s/\ac{GCE}s can be deployed in a distributed manner and organized into multi‑level primary‑secondary cooperation structures. 
    Each \ac{SCE}/\ac{GCE} may directly manage one or multiple \ac{S-gNB}s/\ac{G-gNB}s, whereas a primary \ac{SCE}/\ac{GCE} may also oversee multiple secondary \ac{SCE}s/\ac{GCE}s, thereby indirectly managing one or more \ac{S-gNB}s/\ac{G-gNB}s to improve the efficiency of resource coordination and control.
    \item Unified coordination, configuration and scheduling of satellite-onboard/terrestrial resources can be achieved through the primary \ac{SCE}/\ac{GCE}.
\end{itemize}


\subsubsection{SGCE}
\ac{SGCE} is responsible for cross-domain resource orchestration, global mobility management anchors and long-period policy generation. 
\begin{itemize}
    \item \ac{SGCE} performs resource orchestration and management based on status and service‑demand information reported by \ac{SCE}s and \ac{GCE}s, enabling coordinated decision‑making for multiple integrated space‑ground networks within its jurisdictional region.
    \item \ac{SGCE} can directly manage adjacent \ac{SCE}s/\ac{GCE}s, or indirectly manage other \ac{SCE}s/\ac{GCE}s through adjacent ones, thereby improving the efficiency of resource coordination and control.
    \item Intelligent coordination, configuration and scheduling of communication and computing resources are unified through \ac{SGCE}, enabling highly efficient collaboration between space and ground resources.
\end{itemize}

\subsubsection{\ac{gNB} (\ac{S-gNB} and \ac{G-gNB})}
\ac{gNB} consists of \ac{CU}s, \ac{DU}s and \ac{RU}, adopting a modular functional design pattern to meet the requirements of different application scenarios.
Depending on the different demands of various scenarios, \ac{gNB} can be integrated or separated. 
A single \ac{CU} can manage multiple \ac{DU}s, each of which, in turn, orchestrates one or several \ac{RU}s.

\subsubsection{Signaling interface design between coordinate \ac{NE}s}
By defining dedicated interfaces, the \ac{SGCN} architecture facilitates the efficient exchange and orchestrated management of collaborative information across inter-satellite, inter-terrestrial, and space-ground domains.
\begin{itemize}
    \item \textbf{The \ac{SGCE}-to-\ac{SCE}/\ac{GCE} interface} 
    corresponds to \ac{3GPP} \textit{N2}/\textit{E2} enhancements for cross-domain orchestration,
    facilitating bidirectional interaction and hierarchical management between the centralized \ac{SGCE} and its distributed onboard/terrestrial counterparts (\ac{SCE}s/\ac{GCE}s).
    Specifically, \ac{SGCE} utilizes this interface to disseminate local resource configurations and cross-domain utilization status to \ac{SCE}s/\ac{GCE}s. 
    In return, \ac{SCE}s/\ac{GCE}s feed back their intra-domain resource status and real-time service requirements to \ac{SGCE}, enabling centralized and adaptive resource orchestration.

    \item \textbf{The peer \ac{SCE}/\ac{GCE}-to-\ac{SCE}/\ac{GCE} interfaces}
    facilitate seamless data forwarding within their respective domains. 
    These interfaces can be conceptualized as functional enhancements of \ac{3GPP} \textit{Xn} interface, specifically tailored to support intra-domain mobility and coordinated resource management in the proposed \ac{SGCN}.

    \item \textbf{The \ac{SCE}/\ac{GCE}-to-\ac{gNB} interfaces}
    facilitates local interaction and coordinated management,
    mapping to \ac{3GPP} \textit{F1} enhancements supporting onboard \ac{CU}-\ac{DU} functional splits.
    In an integrated \ac{gNB} configuration,
    \ac{SCE}/\ac{GCE} disseminates local resource configurations, neighboring node utilization profiles (i.e., power allocation, user distribution and spectral interference conditions) to \ac{gNB}. 
    Correspondingly, \ac{gNB} reports its local resource demands and utilization status back to \ac{SCE}/\ac{GCE}. 
    Conversely, in a split-\ac{gNB} architecture, where \ac{CU} and \ac{DU} are separated, \ac{SCE}/\ac{GCE} interacts with \ac{CU} via second-level control data, such as node configuration parameters and cross-node information, allowing \ac{CU} to handle high-layer protocols and non-real-time services. 
    Meanwhile, the interface with \ac{DU} manages millisecond-level telemetry, such as power consumption and the count of scheduled users, enabling \ac{DU} to process low-layer protocols and provide real-time services.
\end{itemize}

\subsection{Analysis of Key Service Capabilities in \ac{SGCN}}
Satellite‑ground coordination achieves large‑scale spatiotemporal collaboration, while air‑interface beam‑level coordination enables small‑scale spatiotemporal collaboration. 
As a whole, they empower the following three primary service capabilities.

\subsubsection{Resource Coordination for Larger Network Capacity}

The proposed \ac{SGCN} achieves resource orchestration through 4D space-ground resource pooling, flexibly allocating frequency, time slots, spatial beams, and power across satellite and terrestrial domains. 
Through coordination between \ac{SGCE} and distributed \ac{SCE}s, cross‑domain resource control resolves significant resource disparities, severe co‑channel interference and complex space-ground resource scheduling. 
Specifically, \ac{SGCE} performs spectrum‑situation prediction and formulates a dynamic multi‑band resource‑sharing strategy between space and ground based on situational awareness and spectrum‑sensing information,
which significantly improves overall resource utilization. 
Meanwhile, \ac{SCE} adaptively coordinates resources to control the interference between space and ground links, thereby improving the \ac{SIR} experienced by users.

The simulation results are shown in Fig.~3a demonstrating that the well-designed \ac{SGCN} architecture successfully addresses the decision-making delay caused by the high-speed movement of \ac{LEO} satellites through resource pooling and intelligent scheduling.
\ac{SGCE} utilizes real-time traffic prediction to detect when terrestrial load approaches a critical threshold and then offload overflow traffic in a elastic way to the on-orbit \ac{SCE} cluster for proximate processing, thereby ensuring service continuity and high reliability.
Compared with typical baseline schemes, the proposed \ac{SGCN} achieves a 20.6~ms ramp latency, which is 16\% lower than B2 (Gateway-only \ac{SDN} coordination) and a 23.6~ms overall peak latency with 94.4\% system reliability. 
The reliability gain over B0 (Terrestrial-only) reaches 49.4 percentage points.

\subsubsection{Access Coordination for Extensive Continuous Coverage}

To address the demand for high‑capacity and rapid access in large‑scale, highly‑dynamic satellite‑ground environments, the \ac{SGCN} architecture supports adaptive access node selection, user contention resolution and seamless cross-domain handover.
Specifically, \ac{SGCE} provides unified decision‑making, 
while distributed \ac{SCE}s implement flexible beam configuration,
enabling joint coverage for terrestrial clustered users and individual users.

Fig.~3b depicts \ac{SGCN}'s capability in managing high-dynamic mobility, compared with three baseline methods.
B0 (Hard handover) suffers an approximately 600~ms interruption at the cell boundary. 
B1 (Conditional handover) reduces peak latency to 38~ms via advance preparation, but retains a 30 ms execution transient.
B2 (Gateway \ac{SDN} Coordination) achieves 26.9~ms peak latency, though mandatory ground \ac{CN} routing adds a persistent penalty.
While the proposed \ac{SGCN} employs trajectory-aware pre-access with \ac{DC} and dynamic weight allocation, achieving a linear and seamless transition of 15.3~ms peak latency—a 39-fold improvement over hard handover.

\subsubsection{Transmission Coordination for Enhanced Spectral Efficiency}


By leveraging the global situational awareness of \ac{SGCE}, the proposed architecture facilitates joint beamforming and \ac{IA} between satellite and terrestrial segments, elevating spectral efficiency by mitigating inter-system interference.
For intra-satellite coordination, \ac{SCE} deploys advanced pre-coding for simultaneous multi-stream transmission.
For multi-satellite coordination, \ac{SCE} exchanges information with \ac{S-gNB} to achieve inter-satellite beam-level time-frequency synchronization, enabling multi-satellite multi-stream data transmission that improves both throughput and spectral efficiency. 
\ac{SGCE} further supports multi-connectivity-based satellite-ground coordinated transmission, dynamically balancing load based on service demands and network capacity.

As illustrated in Fig.~3c, 
a simulation is conducted to evaluate the spectral efficiency of the proposed \ac{SGCN} and three other baselines.
B0 (Single satellite) achieves 12.9~bps/Hz at 20~dB \ac{SNR}, while B1 (Single satellite regenerative) improves to 14.9~bps/Hz through feeder-link penalty elimination. 
B2 (Gateway selection combining) yields only 13.7~bps/Hz, demonstrating that selection combining without coherent combining provides negligible gain. 
The proposed \ac{SGCN} scheme achieves 22.0~bps/Hz—a 71\% improvement—through three synergistic mechanisms, which can be clearly observed from ablation analysis: 
coherent multi-satellite combining adding 4.8~dB effective gain, 
\ac{IA} contributing 2~dB via \ac{SGCE}-coordinated cross-domain beam nulling, 
and \ac{DC} capacity aggregation from the \ac{G-gNB}.

\begin{figure}[!t]
    \centering
    \includegraphics[width=0.495\textwidth]{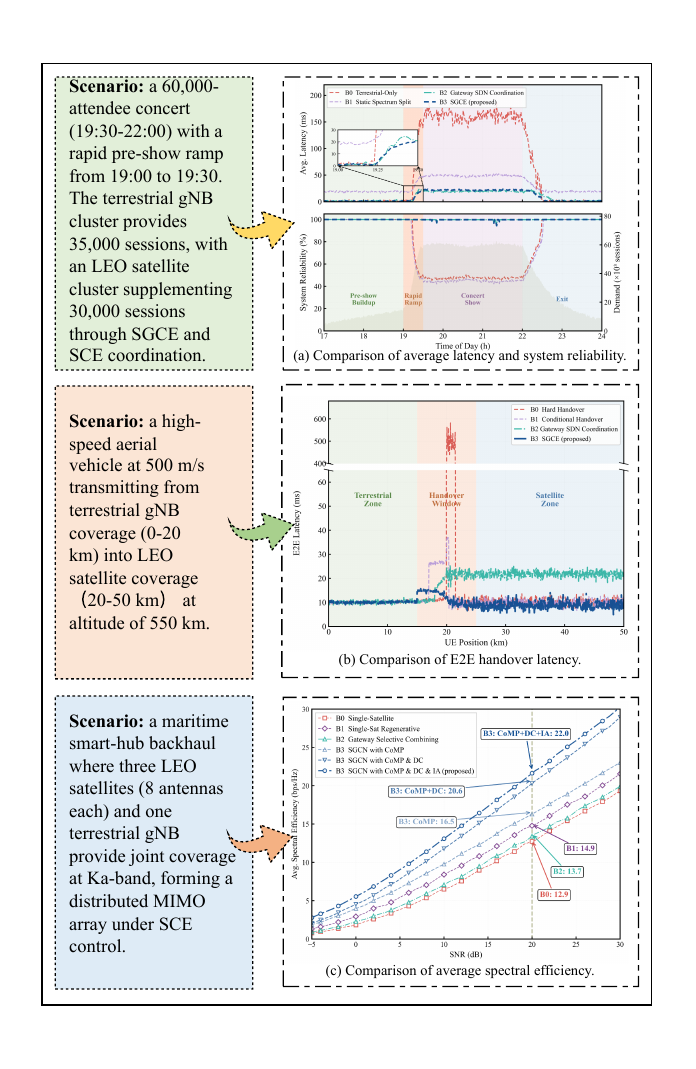}
    \caption{Performance comparisons across three representative scenarios.}
    \label{fig:results}
\end{figure}

\section{Flexible Deployment of Space-Ground Collaborative Network Elements}

\begin{figure*}[tb]
    \centering
    \includegraphics[width=0.96\textwidth]{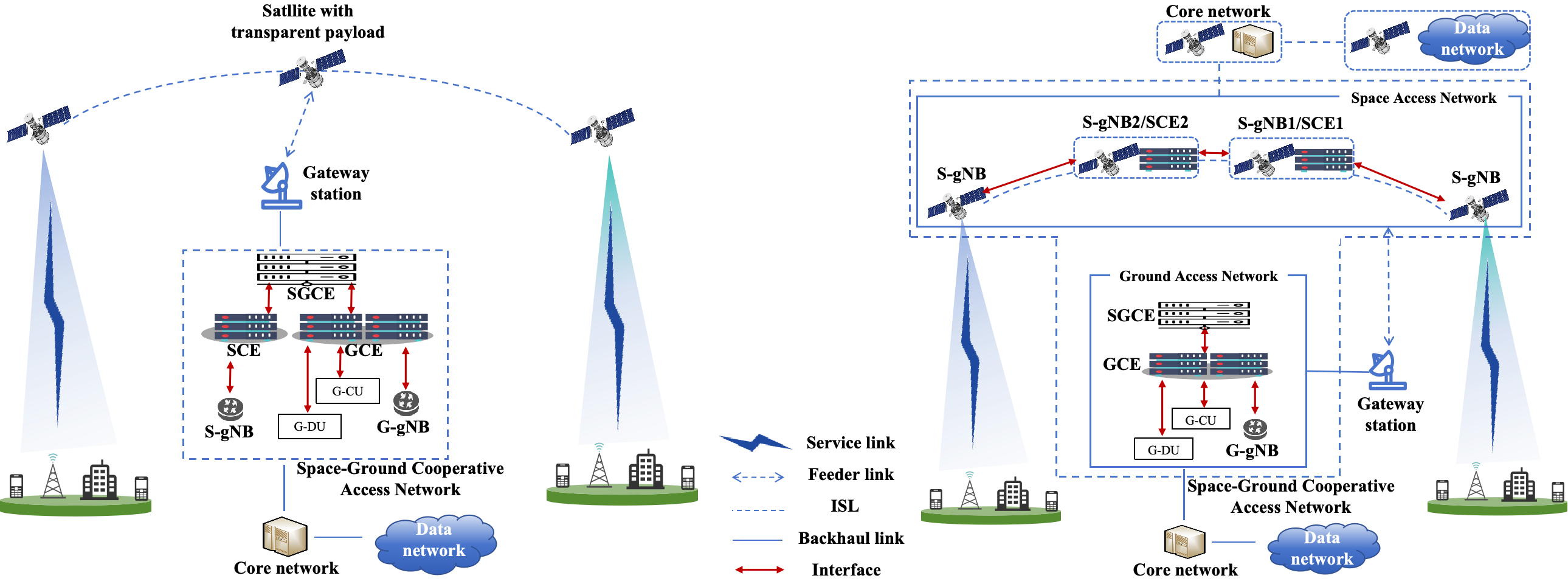}
    \caption{Illustration of two traffic handling methodologies in the proposed \ac{SGCN} based on \ac{NTN}.}
    \label{fig:deployment}
\end{figure*}

A service-aware deployment scheme for collaborative \ac{NE}s is introduced to reconcile the trade-off between robustness and latency.

\subsection{Latency-Tolerant Services}
Latency-tolerant services, exemplified by environmental telemetry and precision agriculture 
are predominantly characterized by \ac{mMTC}, stringent terminal energy constraints and high aggregate data density.
These applications typically utilize the store-and-forward paradigm, wherein satellites function as orbital relays to aggregate data from spatially dispersed sensors for subsequent asynchronous backhauling to terrestrial processing centers.
This operational model not only leverages technical maturity and regulatory stability but also streamlines the seamless migration of legacy \ac{5G} protocol stacks toward integrated \ac{6G} architectures.

However, constrained by \ac{RTT}, it cannot support millisecond-level interactive applications like 
\ac{XR} collaboration. 
Moreover, its performance is sensitive to gateway station density.
e.g., if a station fails or becomes geographically inaccessible,
Therefore, 
a transparent forwarding deployment strategy can be adopted for those latency-tolerant services, as illustrated in the left part of Fig.~\ref{fig:deployment}. 
Within this operational paradigm, satellites equipped with transparent payloads function as bent-pipe relays, tunneling communication signals to terrestrial gateways. 
This traffic is subsequently orchestrated by \ac{SGCE}
to undergo protocol conversion and management before interfacing with \ac{CN} and external data networks.

Consequently, cooperative units can be fully deployed on the ground, with the satellite solely performing transparent data relay, significantly simplifying onboard processing logic and reducing the consumption of onboard resources. 
Under this configuration, ground systems can fully leverage cloud and edge computing resources to enable unified access management, data pre-processing and aggregation for massive \ac{IoT} terminals, thereby enhancing overall network efficiency. 

\subsection{Latency-Critical Services}
For those latency-critical services like emergency communications and autonomous driving assistance, 
there are stringent requirements for latency, bandwidth, and reliability. 
This paradigm shift necessitates the instantiation of collaborative \ac{NE}s within the space-borne segment to facilitate low-latency access control, expedited transmission, and autonomous resource orchestration.
Therefore, it must be deployed on the basis of a regenerative mode, where the satellite possesses onboard processing capabilities. 
Through the collaboration of \ac{SCE} with \ac{S-gNB}s, dynamic inter-beam scheduling, spectrum sharing, and load balancing can be achieved, significantly enhancing the access speed and transmission rates. 

For the satellite regenerative configuration, which is illustrated in the right part of Fig.~\ref{fig:deployment}, the space-based bearer network serves as the core component of the \ac{SGCN}, undertaking distributed switching and intelligent scheduling functions.
To mitigate the constellation's overall \ac{SWaP} overhead, \ac{SCE} is strategically instantiated only on designated master nodes, adhering to a ``one-to-many'' control paradigm where a single \ac{SCE} orchestrates multiple satellites.
\ac{SCE} performs unified resource orchestration, interference coordination and mobility management for its subordinate \ac{S-gNB}s,
and exchanges state information with neighbor \ac{SCE}s to enable cross-satellite joint scheduling and multi-satellite cooperative transmission. 
Furthermore, \ac{SCE} communicates with \ac{SGCE} to coordinate global network slicing, \ac{E2E} \ac{QoS} and spectrum sharing between space and ground segments.

While the regenerative configuration streamlines the user-plane routing and bolsters spectral efficiency, it necessitates more stringent space-borne processing, storage and thermal management capacities. 
Through seamless satellite-terrestrial orchestration, the network facilitates an elastic topology that sustains high-bandwidth, low-latency connectivity even in terrestrial-constrained environments, thereby achieving ubiquitous coverage and significantly bolstering network resilience.

\subsection{Emerging Enhanced Services}
With the development of \ac{6G}, a significant number of hybrid services, such as smart grids and intelligent logistics, which possess both non-real-time collection and real-time triggering capabilities, are rapidly emerging. 

To accommodate these heterogeneous service requirements, \ac{SGCN} employs a dynamic on-demand reconfiguration strategy,
where the transparent mode serves as the default, triggering regenerative activation upon anomaly detection. 
This agile mechanism effectively harmonizes energy conservation during steady-state periods with ultra-reliable performances during transient critical events.

Furthermore, 
the introduction of \ac{AI} agents into decision-making units enables the dynamic and autonomous adjustment of deployment strategies.
As computational entities with sensing, decision-making, and execution capabilities, intelligent agents can continuously acquire real-time information from the network environment, resource status, and task requirements. 
Consequently, decision-making units no longer rely on pre-defined static rules but can autonomously optimize their deployment strategies based on real-time feedback,
This not only enhances the robustness of the system in handling uncertainties but also improves network resource utilization and user experience.

\section{Future Research Trends}
Future research directions for \ac{SGIN} are outlined across the following four dimensions.
\subsection{\ac{AI}-Native and Autonomous Evolution}
Current satellite-terrestrial collaborative decision-making relies heavily on predefined rules and static thresholds, which are inherently limited by the profound spatiotemporal non-stationarity of \ac{LEO} constellation. 

Therefore, future systems must evolve towards an \ac{AI}-native architecture.
Onboard \ac{SCE} inference engines running lightweight quantized models enable real-time traffic prediction and anomaly detection.
The \ac{SGCE} employs large-scale \ac{RL} for global policy optimization as well as constellation topology changes and tidal fluctuations prediction at minute-level granularity, with distilled policies pushed to \ac{SCE}s and \ac{GCE}s for execution.

\subsection{\ac{ISCC}}
Current communication‑centric \ac{SGCN} designs struggle to meet the stringent \ac{QoS} requirements of \ac{6G} services and often suffer from significant performance jitter and beam tracking errors due to a lack of real-time situational awareness.

To overcome these limitations, a deep \ac{ISCC} is urgently required. 
On the one hand, through \ac{SGCE}, computing resources across satellites and ground networks can be uniformly pooled and orchestrated. 
When on‑board computational load becomes excessive, computation tasks can be offloaded via \ac{ISL} to adjacent idle satellites or ground‑based edge nodes.
On the other hand, utilizing the sensing capabilities of satellites enables precise narrow-beam alignment with high-speed terminals, thereby reducing performance fluctuations caused by beam switching.

\subsection{Security and Trust Assurance}
The inherently open nature and global coverage of \ac{LEO} constellations create an expansive attack surface, increasing exposure to illegal eavesdropping and unauthorized access attempts.
On the one hand, it is necessary to investigate physical layer security techniques that leverage the uniqueness of satellite-terrestrial channel characteristics to generate dynamic encryption keys to prevent illegal operations under large-scale coverage scenarios.
However, leveraging the terminal constraints of computing power and energy supply, a lightweight identity authentication protocol across diverse national regulatory domains can be designed to ensure the integrity and privacy of user information during transmission over \ac{ISL}s.

\subsection{High Frequency and Massive \ac{MIMO} Applications}
Attaining \ac{Gbps}-level throughput essential for \ac{6G} broadband services mandates the utilization of high-frequency spectra, specifically \ac{mmWave} and \ac{THz} bands. 
However, signals in these regimes are susceptible to significant atmospheric impairments.
Furthermore, extreme Doppler frequency shifts caused by the high velocity of \ac{LEO} satellites can severely degrade the link budget.

Consequently, it is imperative to engineer 
The distributed orchestration of multiple satellites to synthesize a virtual antenna aperture has emerged as a potential solution for enhancing the uplink link budget. 
This distributed approach, combined with advanced Doppler compensation techniques, is essential for maintaining stable and high-speed connectivity for next-generation mobile applications.

\section{Conclusion}
\label{sec:conclusion}
The transition towards \ac{6G} communication era necessitates a profound architectural evolution from independent satellite and terrestrial networks to a unified, collaborative ecosystem. 
This article proposed a hierarchical \ac{SGCN} architecture,
specifically designed to overcome the inherent barriers of spatiotemporal scale mismatch and extremely high dynamics.

We first summarized the evolution trends of \ac{SGIN}, then categorized the scenarios into two classes and explored some potential technical requirements.
Subsequently, we introduced the design of the hierarchical \ac{SGCN} architecture with explicit \ac{3GPP} standards mapping.
To demonstrate the effectiveness of the proposed architecture, three representative scenarios with typical baselines were simulated and discussed.
Then, a service-aware deployment strategy was proposed, establishing the engineering feasibility and practical deployability of the proposed \ac{SGCN}.
We further identified \ac{AI}-native evolution, \ac{ISCC}, security, and high-frequency technologies as critical enablers of the long-term advancement of \ac{SGCN}.

\section*{Acknowledgement}
This work is supported by National Natural Science Foundation of China under Grant 62401070; Natural Science Foundation of Beijing under Grant L232002; The Open Project of Anhui Provincial Key Laboratory of Multimodal Cognitive Computation, Anhui University, No. MMC202403.

\bibliographystyle{IEEEtran}
\bibliography{reference}

\begin{IEEEbiographynophoto}{Yunxin Geng} (Graduate Student Member, IEEE) received the M.E. degree from Beijing University of Posts and Telecommunications, China, in 2023, where he is currently pursuing his Ph.D. degree. He is an engineer with Shanghai Satellite Network Research Institute Co. Ltd.. His research interests include satellite networks, space-based computing and deep learning.
\end{IEEEbiographynophoto}

\begin{IEEEbiographynophoto}{Shaopeng Feng} received the Ph.D degree from Beijing University of Posts and Telecommunications in 2007. He is currently an Expert with the State Key Laboratory of Satellite Network, Shanghai, China. He has long been engaged in the standard framework design of terrestrial and satellite communications, and holds dozens of invention patents. His research interests include SAGIN architecture and intelligent communications.  
\end{IEEEbiographynophoto}

\begin{IEEEbiographynophoto}{Xuekun Hao} received the Ph.D. degree from the PLA University of Science and Technology, in 2004. He is an expert in satellite communication technologies. He is currently the director with the State Key Laboratory of Satellite Network. His research interests include satellite communication and integrated satellite-terrestrial networks.
\end{IEEEbiographynophoto}

\begin{IEEEbiographynophoto}{Zijian Guo} received the B.S. and M.S. degrees from Beijing Institute of Technology in 2021 and 2024, respectively. He is currently an engineer with Shanghai Satellite Network Research Institute Co. Ltd.. His research interests include integrated space-air-ground networks, satellite internet architecture, and key satellite communication technologies. He focuses on theoretical analysis and engineering studies for future space information networks. 
\end{IEEEbiographynophoto}

\begin{IEEEbiographynophoto}{Shengfeng Wang} received Ph.D. degrees from Beijing University of Posts and Telecommunications (BUPT), Beijing, China. His main research interests include  spacenetwork architecture, complex network dynamics and self-organized system.
\end{IEEEbiographynophoto}

\begin{IEEEbiographynophoto}{Rui Ding} received the B.S. and Ph.D. degrees in electronics and telecommunications from the Southeast University, Nanjing, China, in 2001 and 2011, respectively. He is currently a Professor with China Satellite Network Group Company Ltd., Beijing, China. His research interests include satellite communication and integrated satellite-terrestrial communications networks.
\end{IEEEbiographynophoto}

\begin{IEEEbiographynophoto}{Kan Zheng} (Fellow, IEEE) received the Ph.D. degree from Beijing University of Posts and Telecommunication (BUPT) in 2005. He is currently a full professor in Ningbo University, Zhejiang, China. His research interests include vehicular networks, Internet of Things (IoT) and 6G.  
\end{IEEEbiographynophoto}

\begin{IEEEbiographynophoto}{Zhiqing Wei} (Member, IEEE) received the B.E. and Ph.D. degrees from Beijing University of Posts and Telecommunications (BUPT), Beijing, China, in 2010 and 2015, respectively. He is a Professor
with BUPT. His research interest is the performance analysis and optimization of intelligent machine networks. He was granted the
Exemplary Reviewer of IEEE WIRELESS COMMUNICATIONS LETTERS in 2017, the Best Paper Award of WCSP 2018 and WCSP 2022. He was the Registration Co-Chair of IEEE/CIC ICCC 2018, the publication Co-Chair of IEEE/CIC ICCC 2019 and IEEE/CIC ICCC 2020.
\end{IEEEbiographynophoto}

\begin{IEEEbiographynophoto}{Weiwei Jiang}(Senior Member, IEEE) received the B.Sc. and Ph.D. degrees from the Department of Electronic Engineering, Tsinghua University, Beijing, China, in 2013 and 2018, respectively. He is currently an associate professor with the School of Information and Communication Engineering, Beijing University of Posts and Telecommunications, and Key Laboratory of Universal Wireless Communications, Ministry of Education. His current research interests include artificial intelligence, machine learning, big data, wireless communication and edge computing.
\end{IEEEbiographynophoto}

\begin{IEEEbiographynophoto}{Zhiyong Feng} (Senior Member, IEEE) received the B.E., M.E., and Ph.D. degrees from Beijing University of Posts and Telecommunications (BUPT), Beijing, China. She is a Professor with BUPT, and the Director of the Key Laboratory of the Universal Wireless Communications, Ministry of Education, China. She is a Vice Chair of the Information and Communication Test Committee of the Chinese Institute of Communications (CIC). Her main research interests include wireless network architecture design and radio resource management in 5th generation mobile networks (5G), spectrum sensing and dynamic spectrum management in cognitive wireless networks, and universal signal detection and identification.
\end{IEEEbiographynophoto}

\vfill

\end{document}